\documentclass[amsmath, amsfonts, showpacs, twocolumn, prb]{revtex4}
\usepackage{graphicx}
\usepackage{epsfig}
\usepackage{bm}
\usepackage{dcolumn}
\usepackage{amsmath}
\usepackage{amssymb}
\usepackage{euscript}
\usepackage[varg]{txfonts}

\def\tel{\tau_{\mathrm{el}}}
\def\tgl{\tau_{\mathrm{GL}}}
\def\qo{(q,\omega)}
\def\d{\mathrm{d}}
\def\sgn{\mathrm{sgn}}

\begin{document}

\title{Interaction corrections to tunneling conductance in ballistic superconductors}

\author{Alex Levchenko}

\affiliation{Materials Science Division, Argonne National
Laboratory, Argonne, Illinois 60439, USA}

\begin{abstract}
It is known that in the two-dimensional disordered superconductors
electron-electron interactions in the Cooper channel lead to the
negative logarithmic in temperature correction to the tunneling
conductance, $\delta g_{DOS}\propto-\ln\big(\frac{T_c}{T-T_c}\big)$,
above the critical temperature $T_c$. Physically this result appears
due to the density of states suppression by superconductive
fluctuations near the Fermi level. It is interesting that the other
correction, which accounts for the Maki-Thompson-type interaction of
fluctuations, is positive and exhibits strong power law, $\delta
g_{MT}\propto\big(\frac{T_c}{T-T_c}\big)^3$, which dominates the
logarithmic term in the immediate vicinity of the critical
temperature. An interplay between these two contributions determines
the zero-bias anomaly in fluctuating superconductors. This paper is
devoted to the fate of such interaction corrections in the ballistic
superconductors. It turns out that ballistic dynamic fluctuations
perturb single-particle density of states near the Fermi level at
the energy scale $\epsilon\sim\sqrt{T_c(T-T_c)}$, which is different
from $\epsilon\sim T-T_c$, relevant in the diffusive case. As the
consequence, fluctuation region becomes much broader. In this regime
we confirm that correction to the tunneling conductance remains
negative and logarithmic not too close to the critical temperature
while in the immediate vicinity of the transition we find novel
power law for the Maki-Thompson contribution, $\delta
g_{MT}\propto\big(\frac{T_c}{T-T_c}\big)^{3/2}$. We suggest that
peculiar non-monotonous temperature dependence of the tunneling
conductance may be probed via magneto-tunnel experiments.
\end{abstract}

\date{October 13, 2009}


\maketitle

As it is well known,~\cite{Larkin-Varlamov} the leading-order
fluctuation corrections to the conductivity due to electron-electron
interaction in the Cooper channel in the vicinity of the
superconducting transition are given by the
Aslamazov-Larkin~\cite{Aslamazov-Larkin} (AL),
Maki-Thompson~\cite{Maki,Thompson} (MT), and density of
states~\cite{Dos-diffusive} (DOS) contributions. The first one has a
simple physical meaning of the direct charge transfer mediated by
fluctuation preformed Cooper pairs. The other two contributions have
a purely quantum origin. The Maki-Thompson process can be understood
as the coherent Andreev reflection of electrons on the local
fluctuations of the order parameter while density of states effects
originate from the depletion of energy states near the Fermi level
by superconductive fluctuations. The relative importance of these
three contributions depends on whether a superconductor is diffusive
($T\tel\ll1$), ballistic ($T\tel\gg1$) or granular
($\delta\ll\Gamma\ll\mathrm{max}\{E_{Th},T\}$). Here $\tel$ is the
elastic scattering time on impurities, $\delta$ is the mean level
spacing in the grain, $\Gamma$ is the escape rate, $E_{Th}=D/\ell^2$
is Thouless energy for a grain with the typical size $\ell$, and $D$
is the diffusion coefficient. The Aslamazov-Larkin correction is
essential in both pure and impure superconductors. The Maki-Thompson
is important only in the disordered systems, since there is strong
cancellation between MT and density of states effects in the
ballistic regime.~\cite{Livanov-Varlamov} Usually unimportant DOS
contributions become crucial in the systems containing tunneling
junctions~\cite{Varlamov-Dorin,Tunnel-Exp} or in granular
superconductors.~\cite{SBE,LVV} Tunnel barriers or granularity
require multiple electron scattering for AL and MT contributions to
be important. As the result, the magnitude of these effects is
suppressed either by an extra powers of tunneling matrix element
$\sim|t_{pk}|^2$ (in the case of tunnel
barriers~\cite{Varlamov-Dorin}) or by the small ratio
$g_{\Gamma}/g_{\delta}\ll1$ between inter-grain
$g_{\Gamma}\sim\Gamma/\delta$ and intra-grain $g_{\delta}\sim
E_{Th}/\delta$ conductances (in the case of granular
superconductors~\cite{SBE,LVV}).

In the study~\cite{Varlamov-Dorin} of tunneling anomaly between
diffusive thin-film superconductors separated by an insulating layer
it was shown that there is one specific MT-type process that
contributes significantly to the renormalization of the tunneling
conductance. This process appears to the first order in tunneling
probability $|t_{pk}|^2$, like DOS contribution, however, to the
second order in interaction, unlike DOS, thus containing one extra
power of small Ginzburg number, $Gi\lesssim1$. The reason why these
two contributions have to be accounted simultaneously is twofold.
Unlike DOS correction $\delta g_{DOS}$, which leads to the
suppression of the tunneling conductance above the critical
temperature $T_c$, the novel MT-type contribution $\delta g_{MT}$
leads to its enhancement. Second, although being smaller by the
inverse power of dimensionless conductance $g=\nu
D=k_{F}\ell_{\mathrm{el}}/2\pi\gg1$, where $\nu=m/\pi$ is the
single-particle density of states in two-dimensions and
$\ell_{\mathrm{el}}=v_F\tel$, this specific MT contribution has much
stronger power-law temperature dependence, $\delta g_{MT}\propto
Gi^2\big(\frac{T_{c}}{T-T_c}\big)^3$ , as opposed to the weak
logarithmic in temperature correction coming from the density of
states, $\delta g_{DOS}\propto-Gi\ln\big(\frac{T_c}{T-T_c}\big)$.
One should recall here that the parameter that controls perturbative
expansion over superconductive fluctuations is set by the Ginzburg
number, which is just inversely proportional to the dimensionless
conductance, $Gi\propto1/g$. So that, this is really a competition
between these two contributions that defines the nature of zero-bias
anomaly in fluctuating superconductors. As a result, due to an
opposite signs of $\delta g_{DOS}$ and $\delta g_{MT}$ terms the
full conductance correction $\delta g_{DOS}+\delta g_{MT}$ has
non-monotonous temperature dependence that even may change sign.
Similar observations emerge in the context of granular
superconductors.~\cite{LVV}

The reason for such strong temperature dependence of $\delta g_{MT}$
was attributed in Refs.~\onlinecite{Varlamov-Dorin} and
\onlinecite{LVV} to the importance of vertex renormalization by
coherent impurity scattering (Cooper ladders). If so this would
imply then that anomalous Maki-Thompson contribution is absent in
ballistic tunnel junctions. Such conclusion is also appealing in the
view of strong cancellation of MT effects in ballistic regime of
superconducting thin films.~\cite{Livanov-Varlamov} However, as we
show in this work, in contrast to the expectation MT interaction
correction to conductance in ballistic tunnel junctions remains
important. It also exhibits strong temperature dependence, similar
to that in the diffusive regime, but with the fractional powers
depending on dimensionality.

\begin{figure}
  \includegraphics[width=8.5cm]{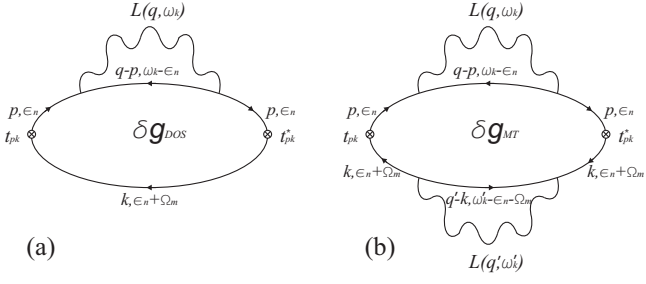}\\
  \caption{DOS and MT tunneling conductance correction diagrams.}\label{Fig-g-DOS-MT}
\end{figure}

In what follows, we carry out a microscopic calculation of
interaction corrections to the tunneling conductance in ballistic
superconductors $T\tel\gg1$ with the help of standard temperature
diagrammatic technique.~\cite{AGD} Within this formalism the
conductance
\begin{equation}\label{g}
g_T=-e\frac{\partial}{\partial
V}\mathrm{Im}\big[\Pi^R(\Omega)\big]_{\Omega=eV}\,,
\end{equation}
is determined by the retarded component of the polarization operator
$\Pi^{R}(\Omega)$. Here $e$ is the electron charge and $V$ is the
voltage applied across the junction. In the case of non-interacting
electrons Matsubara version of the polarization operator is given by
the simple loop diagram, which reads analytically as
$\Pi(\Omega_m)=T\sum_{\epsilon_n}\sum_{pk}|t_{pk}|^2G(p,\epsilon_n+\Omega_m)G(k,\epsilon_n)$,
where $t_{pk}$ stands for the tunneling matrix element,
$\epsilon_n=2\pi T(n+1/2)$ and $\Omega_m=2\pi Tm$ are fermionic and
bosonic Matsubara frequencies, respectively, and
\begin{equation}
G(p,\epsilon_n)=\frac{1}{i\epsilon_n-\xi_p+\frac{i\sgn\epsilon_n}{2\tel}}\,,\quad
\xi_p=\frac{p^2-p^{2}_{F}}{2m}\,,
\end{equation}
defines the single-particle Green's function. Under the assumption
of momentum-independent tunneling amplitudes a simple calculation
then gives for the bare value of the conductance
$g_{T}=\frac{\pi}{2}e^2\nu^2|t_{pk}|^2$. The first-order interaction
correction is given by the diagram shown in Fig.~\ref{Fig-g-DOS-MT}a
\begin{eqnarray}\label{Pi-DOS}
\delta\Pi_{DOS}(\Omega_m)=2T^2\sum_{\epsilon_n\omega_k}\sum_{pkq}|t_{pk}|^2
G^{2}(p,\epsilon_n)\nonumber\\G(k,\epsilon_n+\Omega_m)G(q-p,\omega_k-\epsilon_n)
L(q,\omega_k)\,,
\end{eqnarray}
which amounts an insertion of a single interaction line into one of
the Green's function and coefficient of two accounts for two such
possibilities. This is the density of states effect since upper part
of the diagram is just a self-energy for the $G(p,\epsilon_k)$. The
interaction propagator is defined as
\begin{equation}\label{L}
L(q,\omega_k)=-\frac{8T}{\pi\nu}\frac{1}{Bq^{2}+\tgl^{-1}+|\omega_k|}\,,
\end{equation}
where $ B=\frac{7\zeta(3)v^{2}_{F}}{2d\pi^3T}$ and
$\tgl=\frac{\pi}{8(T-T_c)}$ with $d=1,2,3$ being effective
dimensionality of a superconductor ($1d$ wire, $2d$ film or $3d$
bulk). This approximate form of the interaction is obtained from the
general expression~\cite{Larkin-Varlamov}
$L^{-1}\qo=-\nu\big[\ln\frac{T}{T_c}+\psi\big(\frac{1}{2}+\frac{|\omega_k|}{4\pi
T}\big)+\xi^{2}(T\tel)q^2-\psi\big(\frac{1}{2}\big)\big]$, where
$\xi(T\tel)=\frac{v^{2}_{F}\tel^{2}}{d}\big[\psi\big(\frac{1}{2}\big)+\frac{1}{4\pi
T\tel}\psi'\big(\frac{1}{2}\big)-\psi\big(\frac{1}{2}+\frac{1}{4\pi
T\tel}\big)\big]$, under the assumption that characteristic energies
of fluctuations are small as compared to the temperature,
$\omega\sim T-T_c\ll T$, and momenta are small as compared to the
inverse thermal length $\ell_{T}=v_F/T$, $q\sim\sqrt{1/B\tgl}\ll
\ell^{-1}_T$, which allows to expand digamma functions $\psi$ at
small argument.

Since matrix elements $t_{pk}$ depend weakly on the momenta near the
Fermi-surface one can substitute summation over $p$ and $k$ in
Eq.~\eqref{Pi-DOS} by the corresponding integration over the
energies $\sum_{pk}(\ldots)\Rightarrow\frac{g_{T}}{4\pi
e^2}\int^{+\infty}_{-\infty}\d\xi_p\d\xi_k(\ldots)$. Once these
integrations are performed assuming ballistic limit
$\mathrm{max}\{\epsilon_n,\omega_k\}\gg\tel^{-1}$ and approximating
$\xi_{q-p}\approx\xi_p-v_{F}\cdot q$
\begin{eqnarray}\label{Int-pk}
\sum_{pk}|t_{pk}|^2G^{2}(p,\epsilon_n)G(k,\epsilon_n+\Omega_m)
G(q-p,\omega_k-\epsilon_n)\approx\nonumber\\
-\frac{\pi
g_T}{4e^2}\frac{\sgn(\epsilon_n)\sgn(\epsilon_n+\Omega_m)
\theta(\epsilon_n(\epsilon_n-\omega_k))}
{\big(v_{F}\cdot q+i\omega_k-2i\epsilon_n\big)^2}\,,
\end{eqnarray}
where $\theta(x)$ is the step function, one can complete summation
over the bosonic frequency $\omega_k$ in Eq.~\eqref{Pi-DOS} by
converting it into the contour integral and make an analytical
continuation $i\epsilon_n\to\epsilon+i0$. By combining the result
for $\delta\Pi_{DOS}$ with Eq.~\eqref{g} one obtains density of
states type correction to the zero-bias conductance
\begin{eqnarray}\label{g-DOS}
\hskip-.3cm\frac{\delta
g_{DOS}}{g_T}\!\!&=&\!\!\mathrm{Im}\sum_{q}\int^{+\infty}_{-\infty}
\frac{\d\epsilon}{2T\cosh^2\frac{\epsilon}{2T}}\nonumber\\
&&\int^{+\infty}_{-\infty}\frac{\d\omega}{2\pi}
\frac{L^K\qo+L^R\qo\tanh\frac{\epsilon-\omega}{2T}}
{\big(\omega+v_{F}\cdot q-2\epsilon_+\big)^2}\,,
\end{eqnarray}
where we introduced Keldysh component of the interaction propagator
$L^K\qo=\big[L^R\qo-L^A\qo\big]\coth\frac{\omega}{2T}$ while the
retarded/advances components $L^{R(A)}\qo$ are obtained from
Eq.~\eqref{L} by the replacement $|\omega_k|\to\mp i\omega$. The
most singular in $T-T_c$ contribution to $\delta g_{DOS}$ comes from
the branch-cut of $L^{K}\qo$ where
$L^R\qo\tanh\frac{\epsilon-\omega}{2T}$ term can be ignored. By
taking
$L^K\qo\approx-(32iT^{2}_{c}/\pi\nu)[(Bq^2+\tgl)^2+\omega^2]^{-1}$
one can complete energy integration in Eq.~\eqref{g-DOS} and find
\begin{equation}
\frac{\delta g_{DOS}}{g_T}=\frac{4}{\pi^3\nu}\mathrm{Re}\sum_q
\frac{\psi''\big(\frac{1}{2}+\frac{Bq^2+iv_{F}\cdot
q+\tgl^{-1}}{4\pi T}\big)}{Bq^2+\tgl^{-1}} \,,
\end{equation}
where $\psi''(x)$ is the second derivative of the digamma function.
The remaining $q$ sum is dominated by the small momentum transfer
where argument of the digamma function can be taken as the constant
$\psi''(1/2)=-14\zeta(3)$ since $\mathrm{max}\{Bq^2,v_{F}q,\tgl\}\ll
T$. One obtains then as the result~\cite{Dos-ballistic}
\begin{equation}\label{g-DOS-fin}
\frac{\delta g_{DOS}}{g_T}=-a_d\left\{ \begin{array}{lc}
C_1\sqrt{T_c\tgl} & \quad 1d \\
C_2\ln(T_c\tgl) & \quad 2d
\end{array}
\right.,
\end{equation}
where dimensionless prefactors are $C_1=\frac{1}{\nu
S\sqrt{BT_c}}\propto\frac{1}{p^{2}_{F}S}$ and $C_2=\frac{1}{\nu
Bh}\propto\frac{1}{p_Fh}\frac{T_c}{\epsilon_F}$, with $S$ being the
cross-section area of the wire and $h$ being the thickness of the
film. The numerical coefficients are $a_1\approx2.17$ and
$a_2\approx0.17$. For the bulk $3d$ junctions $\delta
g_{DOS}/g_T\propto-C_3=\frac{T_c}{\nu v_FB}$ is small and
temperature independent. Notice also that for $2d$-case $C_2$ is
linearly proportional to the Ginzburg number. One sees from
Eq.~\eqref{g-DOS-fin} that strong suppression in the density of
states near the Fermi level translates only into moderate
renormalization of conductance $\delta g_{DOS}$. This observation
brings us to the necessity to study contributions to conductance
coming form the interacting fluctuations shown diagrammatically in
Fig.~\ref{Fig-g-DOS-MT}b. The reason for this is similar to that in
the diffusive regime. First of all, this contribution is of the same
order in tunneling $\sim |t_{pk}|^2$ as the density of states one,
Fig.~\ref{Fig-g-DOS-MT}a. Second, although having an extra small
prefactor, $C_d$, this contribution is positive, unlike $\delta
g_{DOS}$, and has much stronger temperature dependence, which may
dominate $\delta g_{DOS}$ in the near vicinity of the critical
temperature. The competition between these terms defines the nature
of zero-bias anomaly in fluctuating regime of ballistic
superconductors.

The diagram in Fig.~\ref{Fig-g-DOS-MT}b defines Maki-Thompson
correction to the polarization operator, which reads explicitly as
\begin{eqnarray}\label{Pi-MT}
\delta\Pi_{MT}(\Omega_m)\!\!=T^3\!\!\!\sum_{\epsilon_n\omega_k\omega'_k}\sum_{pkqq'}
G^2(p,\epsilon_n)G(q-p,\omega_k-\epsilon_n)\nonumber\\
\hskip-.2cmG^2(k,\epsilon_n+\Omega_m)
G(q'-k,\omega'_k-\epsilon_n)L(q,\omega_k)L(q',\omega'_k)\,.
\end{eqnarray}
It is important to comment here that although this correction looks
like second-order DOS, it in fact contains the mixture of advanced
and retarded blocks of the Green's functions, which by its
analytical structure is the same as in the Maki-Thompson diagram.
This is precisely the reason why this term is strongly temperature
dependent. DOS effects always involve Green's functions of the same
causality and thus bring subleading temperature dependence. One
calculates momentum integrals in Eq.~\eqref{Pi-MT} by the
prescription defined in Eq.~\eqref{Int-pk} and after the analytical
continuation finds corresponding correction to the conductance
\begin{eqnarray}\label{g-MT}
\hskip-.5cm&&\frac{\delta
g_{MT}}{g_T}=-\frac{1}{2\pi^2T}\mathrm{Re}\sum_{qq'}
\int^{+\infty}_{-\infty}\frac{\d\epsilon}{\cosh^2\frac{\epsilon}{2T}}
\iint^{+\infty}_{-\infty}\d\omega\d\omega'\nonumber
\\
\hskip-.5cm&&\frac{\mathrm{Im}[L^{R}(q,\omega)]\mathrm{Im}[L^{A}(q',\omega')]
\coth\left(\frac{\omega}{2T}\right)\coth\left(\frac{\omega'}{2T}\right)}
{\big(v_{F}\cdot q+\omega-2\epsilon_{+}\big)^2\big(v_{F}\cdot
q'-\omega'+2\epsilon_{-}\big)^2}\,,
\end{eqnarray}
where $\epsilon_{\pm}=\epsilon\pm i0$ stands as the reminder of
analyticity. After the consecutive energy integrations this formula
simplifies to
\begin{eqnarray}
\frac{\delta
g_{MT}}{g_T}=\frac{256T^{3}_{c}}{\pi\nu^2}\mathrm{Re}\sum_{qq'}
\frac{1}{\big(Bq^2+\tgl^{-1}\big)
\big(Bq'^2+\tgl^{-1}\big)}
\nonumber\\\frac{1}{\big(Bq^2+Bq'^2+iv_{F}\cdot q+iv_{F}\cdot
q'+2\tgl^{-1}\big)^3} \,.
\end{eqnarray}
As compared to the corresponding result in the diffusive
case~\cite{Varlamov-Dorin} the novel feature here is appearance of
the $v_{F}\cdot q$ factors, which limits the phase space for the
momentum transfer and in a way changes power-law behavior of the
singular term in the conductance. The remaining momentum integration
can be completed in the closed form and gives
\begin{equation}\label{g-MT-fin}
\frac{\delta g_{MT}}{g_T}=b_dC^{2}_{d}(T_c\tgl)^{\frac{7-2d}{2}}\,,
\end{equation}
for $d=1,2,3$ with $b_1=0.06$, $b_2=1.4\cdot10^{-3}$, and
$b_3=4\cdot10^{-3}$. As anticipated $\delta g_{MT}$ is positive and
has much stronger power-law temperature dependence then $\delta
g_{DOS}$ near $T_c$. We conclude here that anomalous temperature
dependence of $\delta g_{MT}$ known from the impurity vertex
renormalization in the diffusive case~\cite{Varlamov-Dorin} survives
in the ballistic regime as well, however, with the fractional powers
of $T-T_c$.

\begin{figure}
  \includegraphics[width=8.5cm]{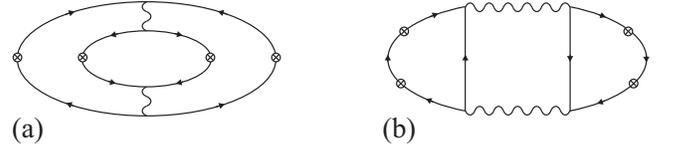}\\
  \caption{Aslamazov-Larkin interaction corrections to the tunneling conductance which
  appear in the higher order in transparency
  $\sim |t_{pk}|^4 $ then DOS and MT contributions shown in Fig.~\ref{Fig-g-DOS-MT}.}\label{Fig-g-AL}
\end{figure}

There are two more diagrams in the second order in interaction that
contribute to the conductance renormalization. These are shown in
Fig.~\ref{Fig-g-AL} and define Aslamazov-Larkin corrections.
However, unlike DOS and MT terms in Fig.~\ref{Fig-g-DOS-MT} these
contributions appear only to the second order in the tunneling
transparency and thus contain and extra smallness $\sim|t_{pk}|^4$.
An estimate for these diagrams for two-dimensional case gives
$\delta g_{AL}/g_{T}\propto|t_{pk}|^2C^{2}_{2}(T_c\tgl)^{1/2}$,
which has smaller amplitude and weaker temperature dependence then
$\delta g_{MT}$ in Eq.~\eqref{g-MT-fin}.

There is simple physical picture which allows to understand these
results at the qualitative level. The current in a tunnel junction
is determined by the product of the density of states convoluted
with the difference of Fermi function, which measure occupation of
the given state, namely $I(V)\sim\int\d\epsilon
[f_{F}(\epsilon+eV)-f_F(\epsilon)]\nu(\epsilon+eV)\nu(\epsilon)$.
Within the linear response one can identify then from $I(V)$ the
zero-bias DOS: $\delta g_{DOS}/g_{T}\sim\int
\frac{\delta\nu(\epsilon)}{\nu}\cosh^{-2}\big(\frac{\epsilon}{2T}\big)\frac{\d\epsilon}{2T}$,
and MT: $\delta g_{MT}/g_{T}\sim\int
\frac{\delta\nu^2(\epsilon)}{\nu^2}\cosh^{-2}\big(\frac{\epsilon}{2T}\big)\frac{\d\epsilon}{2T}$
conductance corrections. Thus, estimation of the temperature
dependence of $\delta g_{DOS}$ and $\delta g_{MT}$ requires
knowledge of the detailed structure of the density of states above
$T_c$. To this end, let us understand at which energy window $\delta
\epsilon$ superconductive fluctuations deplete single-particle
energy states near the Fermi level and what is the depth of this
suppression. The energy scale can be estimated knowing the time
$\tau_\xi$ needed for the superconductive fluctuation to spread over
the distance of coherence length
$\xi(T)=\xi_0\sqrt{\frac{T_c}{T-T_c}}$. In the disordered case
$\tau_{\xi}$ is determined by the diffusive motion of particles and
gives for $\delta \epsilon$ via the uncertainty relation $\delta
\epsilon\sim\tau^{-1}_{\xi}=D\xi^{-2}(T)=\tgl^{-1}\propto T-T_c$. In
the clean limit ballistic motion defines another
scale~\cite{Dos-ballistic} $ \delta
\epsilon\sim\tau_{\xi}=v_F\xi^{-1}(T)\propto\sqrt{T_c(T-T_c)}$. The
depth of the depletion region in DOS,
$\delta\nu(\epsilon)=-\frac{1}{\pi}\mathrm{Im}\Sigma^{R}(\epsilon)$,
follows from the self-energy of the electron Green's function
$\Sigma(\epsilon_n)=T\sum_{pk\omega_n}G^{2}(p,\epsilon_n)G(q-p,\omega_k-\epsilon_n)
\Gamma^{2}(q,\epsilon_n,\omega_k-\epsilon_n)L(q,\omega_k)$, where
impurity vertex
$\Gamma(q,\epsilon_n,\epsilon_m)=\tel^{-1}\theta(-\epsilon_n\epsilon_m)/(Dq^2+|\epsilon_n-\epsilon_m|)$
is present only in the diffusive limit. Having calculated
$\Sigma^{R}(\epsilon)$ DOS renormalization reads
$\delta\nu_{D(B)}(\epsilon)\propto(T_c\tgl)^{\frac{6-d}{2}}F_{D(B)}(2\epsilon\tgl)$,
where $F_{D(B)}(2\epsilon\tgl)$ are energy depending scaling
functions which are universal for the given dimensionality. For
example, in the diffusive $2d$-case~\cite{Dos-diffusive}
\begin{equation}
F_D(z)=\frac{1}{1+z^2}+\frac{(1-z^2)}{2(1+z^2)^2}
\ln\left(\frac{1+z^2}{4}\right)-\frac{2z\arctan(z)}{(1+z^2)^2}\,,
\end{equation}
while in the ballistic regime~\cite{Dos-ballistic}
\begin{equation}
F_{B}(z)=\frac{1}{z^2+\varkappa}
\left[1-\frac{z}{\sqrt{z^2+\varkappa}}
\ln\left(\frac{z+\sqrt{z^2+\varkappa}}{\sqrt{\varkappa}}\right)\right]\,,
\end{equation}
where $\varkappa=(\pi^3/7\zeta(3))T_c\tgl$. The two basic properties
of the scaling functions are $F_{D}(\epsilon\to0)\to\mathrm{const}$
while $F_{B}(\epsilon\to0)\to1/\varkappa$ which is actually valid
for any dimensionality and also $\int^{+\infty}_{-\infty}
F_{D(B)}(\epsilon)\d\epsilon=0$. The latter is the manifestation of
conservation law for the total number of states. Knowing these facts
one readily estimates
$\delta\nu_{D}(0)\propto(T_c\tgl)^{\frac{6-d}{2}}$ and
$\delta\nu_{B}(0)\propto(T_c\tgl)^{\frac{4-d}{2}}$. The most
singular contribution to the MT conductance renormalization comes
from the energy region of maximally depleted $\delta\nu(\epsilon)$
where interaction of superconductive fluctuations is the strongest.
The width of this region is roughly $\delta \epsilon$ and, thus,
interaction correction may be estimated as $\delta
g_{MT}/g_T\propto\delta\nu^2(0)\delta\epsilon$. For the diffusive
case this gives
\begin{equation}
\frac{\delta
g_{MT}}{g_T}\propto(T_{c}\tgl)^{2\times\frac{6-d}{2}}(T_{c}\tgl)^{-1}
\propto\left(\frac{T_c}{T-T_c}\right)^{5-d}\,,
\end{equation}
which reproduces results of Ref.~\onlinecite{Varlamov-Dorin} while
in the ballistic case
\begin{equation}
\frac{\delta
g_{MT}}{g_T}\propto(T_{c}\tgl)^{2\times\frac{4-d}{2}}(T_{c}\tgl)^{-1/2}
\propto\left(\frac{T_c}{T-T_c}\right)^{\frac{7-2d}{2}}\,,
\end{equation}
which agrees with our explicit diagrammatic calculation
[Eqs.~\eqref{Pi-MT}--\eqref{g-MT-fin}]. The reason why $\delta
g_{DOS}$ remains logarithmic in both cases is due to the
conservation law $\int^{+\infty}_{-\infty}
\delta\nu_{D(B)}(\epsilon)\d\epsilon=0$. Indeed, when performing
energy integration in $\delta g_{DOS}/g_{T}\sim\int
\frac{\delta\nu(\epsilon)}{\nu}\cosh^{-2}
\big(\frac{\epsilon}{2T}\big)\frac{\d\epsilon}{2T}$ one necessarily
accounts for the pole of the Fermi function which set the relevant
energies to be of the order of $T$ and not $T-T_c$. For
$\epsilon\sim T$ both scaling functions $F_{D(B)}(\epsilon)$
coincide to the leading singular order in $T-T_c$.

The possible way to probe these temperature anomalies in the
conductance above $T_c$ may be via magneto-tunneling. Let us recall
that magnetic field $H$ acts as an effective Cooper pair breaking
factor that drives a superconductor away from the critical region.
As the result, the relevant energies $\omega$ that determine
conductance corrections in Eqs.~\eqref{g-DOS} and \eqref{g-MT} are
set by the largest cutoff between inverse Ginzburg-Landau time
$\tgl^{-1}\sim T-T_c$ and cyclotron frequency $\omega_H\propto H$,
namely $\omega\sim\mathrm{max}\{\omega_H,\tgl^{-1}\}$. So that by
changing the field one effectively traces temperature dependence of
$\delta g$. Although an explicit calculations of magneto-conductance
in ballistic superconductors is quite involved task we rely here on
the plausible suggestion that is based on the results known for the
diffusive case.~\cite{Reizer} One may expect logarithmic in magnetic
field dependence for the DOS correction $\delta
g_{DOS}(H)/g_T\propto-\ln(T_c/\omega_H)$, when
$\omega_H\gtrsim\tgl^{-1}$, and a power law of $H$ for $\delta
g_{MT}(H)$.

I would like to thank A.~Varlamov for his valuable suggestions and
critical comments that shaped this work and for bringing the
importance of interacting fluctuations in the ballistic limit to my
attention. I am grateful also to M.~Norman, M.~Kharitonov, and
M.~Yu.~Reizer for the useful discussions. This work at ANL was
supported by the U.S. Department of Energy under Contract
No.~DE-AC02-06CH11357.


\end{document}